\documentclass[referee]{aa} 
\usepackage{graphicx}
\usepackage{txfonts}
\def\beq{\begin{equation}}
\def\eeq{\end{equation}}
\def\beqar{\begin{eqnarray}}
\def\eeqar{\end{eqnarray}}

\newcommand{\OOmega}{{\cal A}}

\newcommand{\mN}{{\overline{N}}}

\newcommand{\mrho}{\overline{\rho}}

\newcommand{\trho}{\tilde{\rho}}

\newcommand{\vphi}{\varphi}
\newcommand{\vpsi}{\varpi}
\newcommand{\tn}{\tilde{n}}
\newcommand{\tf}{\tilde{f}}
\newcommand{\thh}{\tilde{h}}
\newcommand{\hh}{\hat{h}}

\newcommand{\hrho}{\hat{\rho}}

\newcommand{\bu}{{\bf u}}
\newcommand{\bbv}{{\bf v}}

\newcommand{\bk}{{\bf k}}
\newcommand{\p}{{\partial}}

\newcommand{\tv}{{\tilde { v}}}
\newcommand{\hf}{{\hat{ f}}}
\newcommand{\hv}{{\hat{ v}}}

\newcommand{\hhf}{{\hat{\hat{h}}}}
\newcommand{\hhv}{{\hat{\hv}}}

\newcommand{\hhh}{\hat{\hh}}
\newcommand{\hhrho}{\hat{\hrho}}

\newcommand{\gapprox}{\lower.4ex\hbox{$\;\buildrel
>\over{\scriptstyle\sim}\;$}}
\newcommand{\lapprox}{\lower.4ex\hbox{$\;\buildrel
<\over{\scriptstyle\sim}\;$}}

\begin{document}
   \title{Self-consistent theory of turbulent transport 
in the solar tachocline}

        \subtitle{III. Gravity waves}

   \author{Eun-jin Kim and Nicolas Leprovost
          }

   \offprints{E. Kim}

   \institute{Department of Applied Mathematics, University of Sheffield,
              Sheffield, S3 7RH, UK\\
              \email{e.kim@sheffield.ac.uk}
             }

   \date{Received 5 July, accepted, 2006}

  \abstract
     {}
   {To understand the fundamental physical processes important for
   the evolution of solar rotation and distribution of chemical 
   species, we provide theoretical predictions for particle
    mixing and momentum transport in the stably stratified tachocline.}
{By envisioning that turbulence is driven in the tachocline,
we compute the amplitude of turbulent flow, 
turbulent particle diffusivities, and eddy viscosity, by incorporating
the effect of a strong radial differential rotation and stable
stratification. We identify the different roles that the 
shear flow and stable stratification play in turbulence regulation
and transport.}
   {Particle transport is found to be severely quenched 
   due to stable stratification as well as radial differential
   rotation, especially in the radial direction with an effectively
   more efficient horizontal transport. The eddy viscosity is
  shown to become negative for parameter values typical of the
tachocline, suggesting that turbulence in the stably stratified
tachocline leads to a non-uniform radial differential rotation.
Similar results also hold in the radiative interiors of stars,
in general.} 
   {}
\keywords{Turbulence -- Sun: interior -- Sun: rotation -- waves }                                                                                                                                                                   
\titlerunning{Gravity waves} 

   \maketitle


\section{Introduction}

Since the formation of the radiative core, which marked the beginning of
its journey on the main sequence, the sun has slowed down significantly
due to the loss of angular momentum from its surface (e.g. see Stix 1989;
Schatzman 1993). 
The angular
momentum transport must have been very efficient during its spin-down
in order for the sun to have rotational profile as
observed today (see, e.g. Charbonneau et al 1998).
Vigorous turbulence in the convection zone 
and possibly thermal wind (Miesch, Brun \& Toomre 2006)
can readily provide
a mechanism for efficient radial momentum transport, thereby eradicating
radial differential rotation therein. Such turbulent transport
is however considered to be absent in the stably stratified
radiative interior, which has also spun-down during the solar evolution,
presently rotating uniformly at the rate roughly the same as the 
mean average rotation rate on the solar surface.
Whichever mechanism is responsible
for momentum transport in the interior (which itself is an
important problem), it should be closely related
to the transport in the tachocline through which the surface spin-down
is communicated to the interior. Transport in the tachocline also plays a 
crucial
role in the overall mixing of light elements (lithium, beryllium, etc)
(see e.g. Schatzman 1993; 
Brun, Turck-Chi\'eze, \& Zahn 1999),
thereby determining the level
of their surface abundances on the sun. 
Therefore, it is essential to understand physical
mechanisms for transport in the tachocline and then to formulate
a consistent theory starting from first principles based on
those processes. This is particularly
true since virtually all the previous theoretical modelling heavily
relies on a simple parameterization of transport process, which is then adjusted
to obtain the agreement with observations.  

We have initiated the development of
a consistent theory of turbulent transport
in the tachocline in the previous papers, by taking into account the 
crucial effect of the large-scale shear flows, provided by a strong radial
differential rotation (Kim 2005) as well as latitudinal
differential rotation (Leprovost \& Kim 2006).
By envisioning that the tachocline is 
perturbed externally
[e.g. by plumes penetrating from the convection zone above
(e.g. see Gilman 2000; Brummell, Clune \& Toomre 2002; 
Rogers \& Glazmaier 2005)], or by instabilities
(e.g. see Watson 1981;
Charbonneau, Dikpati \& Gilman 1999),
we demonstrated how turbulence level and transport
are reduced via shearing in a non-trivial manner 
(see also Burrell 1997; Kim \& Diamond 2003; Kim 2004;2006)
in a simplified three dimensional (3D) hydrodynamic turbulence. 
In particular, turbulent transport
of chemical species and angular momentum are shown to become
strongly anisotropic
with effectively much more efficient transport in the horizontal (latitudinal)
direction than the vertical (radial) direction due to shear stabilization
by strong radial shear.
The resulting anisotropic momentum transport was shown
to reinforce a strong radial shear (Leprovost \& Kim 2006), 
with a positive feedback on the confinement of the tachocline (Spiegel \& Zahn 1992)
while chemical species are predicted to have 
latitudinal
dependent mixing due to the variation of radial shear (Kim 2005).
Furthermore, the results indicate that the turbulence regulation
by a shear flow (i.e. differential rotation) leads to weak
turbulence and mixing in turbulent tachocline.

The purpose of
this paper is to investigate how a stable stratification 
in the tachocline modifies the predictions obtained in these studies.
We again envision that the tachocline is turbulent, 
driven by a forcing as in Kim (2005).
In the presence of a stable
stratification, the turbulence in the tachocline is no longer
completely random as the stable stratification provides a restoring 
force against radial displacement
of fluid elements, supporting the propagation of internal gravity waves
(Lighthill 1978).
These waves tend to increase the memory of otherwise random
turbulent fluid motion and can reduce the overall transport
due to turbulence. We shall show that the turbulent transport due to shear
stabilization found in the previous
studies is further enhanced in the presence
of stable stratification (gravity waves).
Thus, gravity waves acting together with shear stabilization 
can lead to a weak mixing in the tachocline,
as required for the
surface depletion of lithiums (e.g. Pinsonneault
et al.  1989).

It is important to contrast our approach to those adopted in most
previous works, which focus
on the momentum transport by gravity waves themselves through 
dissipative processes (e.g. Plumb 1978;
Kim and MacGregor 2001;2003;
Talon, Kumar \& Zahn 2002). For instance, gravity waves are considered
to be generated in the convection zone (Press 1981)
and deposit their momentum
into background shear flow via radiative damping as
they propagate through the tachocline
and the interior (Kim and MacGregor 2001;2003).
Relying crucially on radiative damping, the momentum transport
by gravity waves would occur on a long time scale. [Recall that
the transport by waves requires molecular dissipation,
and is thus a slow process.] Similarly, weak mixing due to 
damped gravity waves was also suggested as a mechanism for
a modestly enhanced mixing of light elements (lithium) (Press \& Rybicki
1981; Garc\'ia L\'opez \& Spruit 1991).
In these works, a certain level of background turbulence
could not be ruled out and was invoked to provide
an enhanced viscosity for the evolution
of a mean flow (Kim and MacGregor 2001;2003) and the 
gravity waves (Charbonnel \& Talon 2005; 
Talon, Kumar \& Zahn 2002). The value of the
effective viscosity is often arbitrarily taken to be
a positive constant, being much larger than molecular
viscosity, 
not affected by shear flow nor by gravity waves,
or it is simply parameterized.

In this paper, we shall treat turbulence and gravity waves
on an equal footing and consistently 
compute the values of turbulent eddy viscosity and particle
diffusivity, by incorporating
the effect of a shear flow (provided by a radial differential
rotation), and 
identify different roles that gravity waves and shear flow
play in  turbulent transport.
Specifically, we shall show that unlike shear flows, which reduce both
turbulence level and transport, a stable stratification can
suppresses turbulent transport without much effect on turbulence level. 
We shall further demonstrate that the stratification favors a negative
eddy viscosity and thus tends to sharpen
the radial gradient of large-scale shear flows rather than smoothing it out.
This tendency was also found in Kim \& MacGregor (2001;2003).
We note that the elucidation of the effects of gravity waves is essential
for understanding the momentum transport in the radiative interior
as they have often been advocated as a mechanism to explain
a uniform rotation in that region. 
Stratified turbulence with a shear flow is also important 
for the transport in radiative interiors and/or envelops of 
stars, in general, as well as in geophysical systems.
In particular, it has actively been
studied in geophysical
systems (e.g. see Jacobitz, Sarkar \& van Atta 1997; 
Stacey, Monismith, \& Burau 1999), 
where stable stratification was shown to inhibit 
turbulent transport in the direction of a background density gradient,
leading to the two dimensional (2D) turbulence.

The remainder of the paper is structured as follows. 
We first investigate the effect of gravity waves on turbulence
in Sect. 2. In Sect. 3, we incorporate the effect of strong
radial differential rotation and study how the gravity
waves modify the overall turbulent transport.  
Section 4 contains the
conclusion and discussions.

\section{Internal Gravity Waves}

\begin{figure}[h] 
\begin{center} 
\includegraphics[scale=0.9,clip]{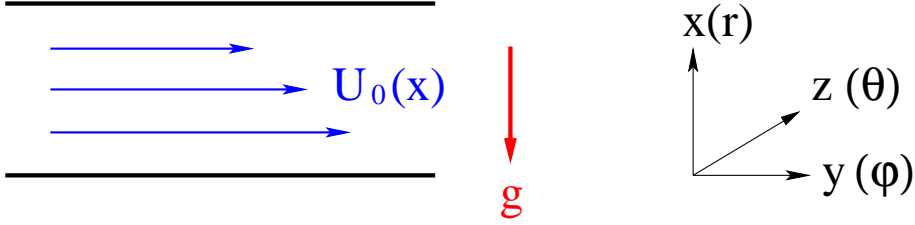} 
\end{center} 
\caption{\label{DessinEq} The configuration of our model.}
\end{figure} 

To simplify the problem, we shall consider 
incompressible fluid with local cartesian coordinates
$x, y$, and $z$ for 
radial, azimuthal, and latitudinal directions, respectively
(see Fig. 1),
and use Boussinesq approximation to capture the effect of
stratification. 
We assume that the fluid is stirred by a forcing 
on small scales, giving rise to fluctuations.
In the absence of stratification, this fluid 
forcing will drive turbulence on small scales
and maintain it at the level at which the injected energy is
balanced by dissipation in the system. 
As the stratification increases,
the forcing will generate not only
random turbulent motion but also coherent (gravity) waves.
Alternatively, some of turbulent motion will be turned into
packets of gravity waves. 
Therefore, we consider both turbulence and gravity waves 
as fluctuations on scales much smaller than those
associated with mean density or mean background shear flows.
Specifically, we express the total mass density
$\rho=\rho_0({x})+\rho'$ where $\rho_0({x})$ and $\rho'$ are
the background and fluctuating mass densities, respectively;
the total velocity ${\bu} = {\bf U}_0 + {\bf v}$ where
${\bf U}_0$ and ${\bf v}$ are a large-scale shear flow 
due to differential rotation and small-scale fluctuations;
the total particle density of chemical elements $n = n_0({\bf x})
+ n'$ where $n_0({\bf x})$ and $n'$ are mean and
fluctuating components.
Then, the main governing equations for fluctuations ${\bf v}$, $\rho'$, 
and $n'$, involving both turbulence and gravity waves, are as follows 
(see, e.g., Kim \& MacGregor 2003; Moffatt 1978): 
\begin{eqnarray}
(\p_t  + {\bf U}_0 \cdot \nabla) \bbv 
&=& -\nabla p -g \rho' { \hat x}+ \nu \nabla^2 \bbv + {\bf f}\,,
\label{eq1}\\
\nabla \cdot \bbv &=&  0\,,
\label{eq2}\\
(\p_t + {\bf U}_0 \cdot \nabla) \rho' 
&=& {{{\overline \rho}} N^2 \over g} v_{x}+ \mu \nabla^2 \rho'\,,
\label{eq3}\\
(\p_t + {\bf U}_0 \cdot \nabla) n' 
&=& -{\p_x n_0} v_{x}+ D \nabla^2 n'\,.
\label{eq4}
\end{eqnarray}
Here,
$\nu$, $\mu$ and $D$ are molecular viscosity, thermal diffusivity,
and particle diffusivity, respectively;
${\bf f}$ in Eq. (\ref{eq1}) is the small-scale forcing
driving turbulence;
${\mrho}=\rho_0(x=0)$ is the constant background density
[measured at the bottom of the convection zone (e.g. see
Kim and MacGregor 2003)],
and 
$N=\sqrt{-g(\p_x \rho_0 + {\mrho} g/c_s^2)/ \overline {\rho}}$ 
is the Brunt-V\"ais\"al\"a frequency, where
$c_s$ is the sound speed.
Note that the typical values of 
$\nu$, $D$, $\mu$, and $N$ in the tachocline are
$10^2$ cm$^2$s$^{{-1}}$,$10^2$ cm$^2$s$^{{-1}}$, 
$10^7$ cm$^2$s$^{{-1}}$, and $3 \times 10^{-3}$ s$^{-1}$, 
respectively.

In this section, we first ignore a background shear flow
and study how the turbulence and transport 
are modified due to stable stratification in the tachocline.
To obtain the overall turbulent transport,
we solve coupled equations (\ref{eq1})-(\ref{eq4})
in terms of Fourier transform for fluctuating quantities $\phi'$:
\begin{equation}
\phi'({\bf x},t) = {1\over (2 \pi)^3} \int d^3 k\,
\tilde{\phi}({\bf k},t) \exp{\{i(k_x x + k_y y + k_z z)\}}\,.
\label{eq09}
\end{equation}
Equations (\ref{eq1})-(\ref{eq3}) then give us an equation for
${\tilde{\trho}}=e^{\nu k^2 t}\trho$ in the form
\begin{equation}
\p_{tt} {\tilde{\trho}} + (\mu - \nu) k^2 \tilde{\trho} + \mN^2 \tilde{\trho}
= {N^2 \over g} e^{\nu k^2 t} \thh_1\,.
\label{eqrho}
\end{equation}
Here, ${\mN}^2=\gamma N^2/(\gamma + a^2)$.
Since $N \gg \mu k^2$ is valid 
on a broad range of reasonable scales $l\gg 10^{4} \sim 10^5$ cm
in the tachocline, we can find the solutions to leading order in $(\nu k^2/N \ll)\,\, \mu k^2/N \ll 1$ as: 
\begin{eqnarray}
{\tv}_x &\sim  &{1\over \gamma + a^2}
\int dt'  G_\mu(t,t') \cos{{\mN(t-t')}}{\thh}_1 (t')\,,
\label{eq01}\\
{\tv}_y &\sim  &- {a\over \gamma} {\tv}_x
- {\beta \over \gamma} 
\int dt'  G(t,t') {\thh}_2 (t')\,,
\label{eq02}\\
{\tv}_z &\sim & - {a\beta \over \gamma} {\tv}_x
+ {1\over \gamma } 
\int dt'  G(t,t') {\thh}_2 (t')\,,
\label{eq03}\\
{\trho} &\sim  &{N\over g \sqrt{\gamma(\gamma + a^2)}}
\int dt'  G_\mu(t,t') \sin{{\mN(t-t')}}{\thh}_1 (t')\,.
\label{eq04}
\end{eqnarray}
Here, $G(t,t') = e^{-\nu k^2(t-t')}$ and
$G_\mu(t,t') = e^{-(\mu+\nu) k^2(t-t')/2}$ 
are the
Green's functions;
$a=k_x/k_y$, $\beta=k_z/k_y$,
and $\gamma = 1+\beta^2$;
$\thh_1$ and $\thh_2$ in Eqs. (\ref{eqrho}) and 
(\ref{eq01})-(\ref{eq04}) are the forcing terms which are related to
${\tilde f}_i$ as 
\begin{eqnarray}
\thh_1 &=& (1+\beta^2) \tf_x - a\tf_y - a \beta \tf_z\,,
\label{eq16}\\
\thh_2 &=& - \beta \tf_y + \tf_z\,.
\label{eq17}
\end{eqnarray}
Equations (\ref{eq01})-(\ref{eq03}) show that the vertical motion
$\tv_x$, involving $G_\mu$ only, is always subject to radiative 
damping $\mu$ while the horizontal motions $\tv_y$ and $\tv_x$, 
containing both $G_\mu$ and ${G}$, are much
less affected by $\mu$.

For simplicity, we assume the forcing to be homogeneous in space with
a short correlation time $\tau_f$:
\begin{equation}
\langle {\tilde f}_i (\bk_1,t_1) {\tilde f}_j (\bk_2,t_2) \rangle
= \tau_f (2 \pi)^3 \delta (t_1-t_2) \delta(\bk_1+\bk_2)
{\psi}_{ij} (\bk_2)\,.
\label{eq18}
\end{equation}
Note that in the case where the forcing ${\bf f}$ due to plumes induces
a stronger turbulence towards the bottom of the convection zone,
${\bf f}$ becomes inhomogeneous with the power spectrum
${\psi}_{ij} = \psi_{ij}(x, {\bf k})$ depending on 
the radial coordinate $x$.

A long but straightforward algebra by using
Eqs.\ (\ref{eq01})--(\ref{eq18}) 
then gives us the following results on turbulence level
and particle diffusivities defined by
$\langle n' v_i \rangle = -D_T^{ij} \p_j n_0$:

\begin{eqnarray}
\langle v_x^2 \rangle
&\sim& {\tau_f} \int {d^3k \over (2 \pi)^3}
{\phi_{11}(\bk)\over 2 (\gamma+a^2)^2}
{1 \over (\mu +\nu) k^2}\,,
\label{eq19}\\
\langle v_y^2 \rangle
&\sim& {\tau_f} \int {d^3k \over (2 \pi)^3}
\left[{a^2 \phi_{11}(\bk)\over 2 \gamma^2 (\gamma+a^2)^2}
{1 \over (\mu +\nu) k^2}
+{\beta^2 \phi_{22}(\bk)\over  \gamma^2 }
{1 \over 2 \nu k^2} \right]\,,
\label{eq20}\\
\langle v_z^2 \rangle
&\sim& {\tau_f} \int {d^3k \over (2 \pi)^3}
\left[{a^2 \beta^2 \phi_{11}(\bk)\over 2 \gamma^2 (\gamma+a^2)^2}
{1 \over (\mu +\nu) k^2}
+{\phi_{22}(\bk)\over  \gamma^2 } 
{1 \over 2 \nu k^2} \right]\,,
\label{eq21}\\
D_T^{xx}
&\sim& {\tau_f\over  N^2} \int {d^3k \over (2 \pi)^3}
{\phi_{11}(\bk)\over 4 \gamma (\gamma+a^2)}\,,
\label{eq019}\\
D_T^{yy}
&\sim & {\tau_f} \int {d^3k \over (2 \pi)^3\gamma^2}
\left[{a^2 \phi_{11}(\bk)\over  4 \mN^2 (\gamma+a^2)^2}
+{\beta^2 \phi_{22}(\bk) 
\over 2\nu (\nu + D)^2 k^4} \right]\,,
\label{eq020}\\
D_T^{zz}
&\sim & {\tau_f} \int {d^3k \over (2 \pi)^3\gamma^2}
\left[{a^2 \beta^2 \phi_{11}(\bk)\over 4 \mN^2 (\gamma+a^2)^2}
+{\phi_{22}(\bk)\over 2\nu (\nu + D)^2 k^4} \right]\,.
\label{eq021}
\end{eqnarray}
Here, we assumed 
spatial symmetries $\phi_{ij} (k_y) = \phi_{ij}(-k_y)$
and $\phi_{ij} (k_y) = \phi_{ij}(-k_y)$;
${\bf k}$ inside the integrals in
Eqs. (\ref{eq19})--(\ref{eq021}) is the wavenumber of
the forcing;
$\phi_{ij}(\bk)$ ($i=1,2$) is the power spectrum of
the forcing defined by:
\begin{equation}
\langle {\tilde h}_i (\bk_1,t_1) {\tilde h}_j (\bk_2,t_2) \rangle
= \tau_f (2 \pi)^3 \delta(t_1-t_2) \delta(\bk_1+\bk_2)
{\phi}_{ij} (\bk_2)\,.
\label{forcing}
\end{equation} 
In this paper, we shall consider an incompressible forcing
($\nabla \cdot {\bf f}=0$) for simplicity, in which case
$\phi_{ij}$ in Eq. (\ref{forcing}) is related to  $\psi_{ij}$ 
in Eq. (\ref{eq18}) as:
\begin{eqnarray}
\phi_{11} &= &{ k^4\over k_y^4} \psi_{11}\,,
\phi_{12} = {1\over k_y^4} ( k^2 k_x k_z \psi_{11}
+ k^2 k_H^2 \psi_{13})\,,
\nonumber\\
\phi_{22} &= &{1\over k_y^4} 
( k^2_x k_z^2 \psi_{11} + k_H^4 \psi_{33} + 2 k_x k_z k_H^2 \psi_{13})\,.
\label{eq22}
\end{eqnarray}
Note that $\phi_{11}$ and $\phi_{22}$ can signify
strong radial forcing (e.g. by plumes) and horizontal forcing
[e.g. due to the instability of the latitudinal shear in the
tachocline (see also Kim 2005)].
For instance,
in the case of strong radial forcing by plumes
($\psi_{11}\gg \psi_{33}$) with $k_H\gg k_x$ (see Fig. 2 in Leprovost
\& Kim 2006), 
$\phi_{11}$ will dominate over $\phi_{22}$.
Note further that in the 2D limit with $k_z=0$ and $\psi_{33} =0$,
$\phi_{22}$ vanishes.

We discuss some of the important implications of Eqs. (\ref{eq19})--(\ref{eq021}).
First, all three components of fluctuating velocity amplitude 
in Eqs. (\ref{eq19})-(\ref{eq21}) are independent of $N$, clearly showing 
that the amplitude of the turbulent flow is not influenced by stable 
stratification
{in the case of the forcing with a short correlation time.}
The exact level of
fluctuations is determined by the characteristics of the external forcing 
($\phi_{ij}$). We examine this in the simple 2D limit where $k_z = 0$
($\gamma = 1$) and $\psi_{33}=0$.
In this limit, the substitution of $\phi_{22}=0$ 
in Eqs. (\ref{eq19})--(\ref{eq21})
gives $\langle v_x^2 \rangle/\langle v_y^2 \rangle \sim 1/a^2
= k_y^2/k_x^2$ and $\langle v_z^2 \rangle = 0$. 
This is an expected result for the 2D incompressible fluid
in $x-y$ domain. If the forcing ${\bf f}$
contains only gravity modes, the wave number of the forcing
would satisfy the local dispersion
relation $k_x = \pm k_y \sqrt{N^2/(\omega-U_0 k_y)^2-1}$
(here, $\omega$ is the frequency of the gravity waves).
Furthermore, if these gravity modes are generated by the overturning of
fluids in the convection zone (e.g. Press 1981) with $U_0\sim 0$, 
they would have a strong power
at low frequencies $\omega \ll N$, thereby giving 
$k_x/k_y \sim N/\omega \gg 1$, and thus 
$\langle v_x^2 \rangle/\langle v_y^2 \rangle \ll 1$.
Note that this is the situation normally considered in the previous
works (e.g. Kim \& MacGregor 2001;2003; Talon, Kumar \& Zahn 2002), 
where the main focus was on the momentum deposition by such gravity waves 
due to radiative damping after entering the tachocline from
the bottom of the convection zone. 
In contrast, in this paper, the forcing ${\bf f}$ is not restricted 
to gravity modes, but is taken to be general, including
any form of perturbation with arbitrary values of $k_x/k_y$. 
For instance, in the case of
a strong radial forcing due to plumes with $\phi_{11} \gg \phi_{22}$
and $a=k_x/k_y \ll 1$ (see Fig 2 in Leprovost \& Kim 2006),
$\langle v_x^2 \rangle/\langle v_y^2 \rangle \gg 1$
with a stronger radial fluctuation than horizontal one. 
Furthermore, Eqs. (\ref{eq19})-(\ref{eq21}) show that
the level of anisotropy measured by the
ratio of the turbulence amplitude depends on $\nu$ and $\mu$.
For instance, for an isotropic forcing with $\phi_{11}\sim {\phi}_{22}$,
$\langle v_x^2 \rangle/\langle v_y^2 \rangle \propto \nu/\mu \ll 1$
becomes small as the radiative damping $\mu$ increases.
This is because a large $\mu$ tends to decrease (vertical) turbulence
level by introducing large (thermal) dissipation. 
Thus, without a shear flow, a stronger horizontal turbulence 
(level) than vertical one can be caused by large thermal diffusivity $\mu$
(but not by stratification) in the case of a temporally short correlation
forcing.
Note that the anisotropy in turbulence level could be related
to the Peclet number $Pe = v l/\mu$ as 
$\langle v_x^2 \rangle/\langle v_y^2 \rangle \propto Pe$.
Here, $v$ and $l$
are the characteristic velocity and length scale of the forcing (which is
fixed).

Second, turbulent transport in Eqs. (\ref{eq019})--(\ref{eq021})
are strongly affected by stratification as $N^2$ increases,
in contrast to fluctuation levels in Eqs. (\ref{eq19})--(\ref{eq21}).
Since 
$N \gg \mu k^2$ on reasonable scales $l\gg 10^{4} \sim 10^5$ cm,
the vertical (radial) mixing is severely quenched
as the stratification increases (i.e. as $N^{-2}$),
in qualitative agreement with Brun, Turck-chi\'eze \& Zahn (1999).
In comparison, the part of the horizontal (latitudinal) mixing 
due to the radial forcing $\phi_{11}$ is reduced
proportional to $N^{-2}$ while the one due to the horizontal forcing 
$\phi_{22}$ is independent
of $N$. The comparison of these two contributions (or alternatively
$D_{xx}$ and $D_{yy}$) in the
case of $\phi_{11} \sim \phi_{22}$ gives us
a cut-off scale $l_c \sim \sqrt{\nu/N}$ 
above which vertical (radial) mixing is strongly reduced
compared to the horizontal (latitudinal) mixing.
That is, in the presence of both radial and horizontal
forcings of comparable strength,
stable stratification mainly reduces the vertical transport
without much effect on horizontal transport on 
scales $l>l_c$. For parameter values typical of the tachocline
$\nu \sim 10^2$ cm$^2$s$^{-1}$ and $N \sim 3 \times 10^{-3}$s$^{-1}$,
$l_c \sim 10^{2}$cm.
Thus, the stratification is very likely
to play an important role over a broad range of physically reasonable
scales. In the limit of strong radial forcing with $\phi_{11}
\gg \phi_{22}$ and $a \sim 0$, the stratification influences both
the radial and horizontal transports to the same degree
while the incompressibility renders $D_T^{xx}/ D_T^{yy} \sim D_T^{xx}/D_T^{zz}
\sim 1/a^2 \sim \langle v_x^2 \rangle / \langle v_y^2 \rangle \gg 1$, 
with an effectively more efficient radial
transport.
It is interesting to compare our result $D_{xx} \propto N^{-2}$
with the mixing due to radiatively damped waves 
(e.g. Garc\'ia L\'opez and Spruit 1991; Talon et al 2002). 
For instance,
Garc\'ia L\'opez and Spruit (1991)
estimated the vertical mixing
due to gravity waves to be proportional to $\mu/N^2$. While the reduction
in vertical mixing for large $N$ is in qualitative agreement
with our results ($\propto N^{-2}$), the increase in vertical mixing
in Garc\'ia L\'opez and Spruit (1991)
is due to the fact that damped waves are necessary for wave transport.
Finally, we note that 
without a shear flow, momentum transport vanishes
(i.e. $\langle v_x v_y \rangle = \langle v_x v_z \rangle = 0$)
for an isotropic forcing.
Scaling of turbulence amplitude, 
turbulent viscosity ($\nu_T$), and turbulent diffusivity $D_T$ 
are summarised in Table 1
for an isotropic forcing.

A large thermal diffusivity ($\mu = 10^5 \nu$) in the tachocline
is often considered to reduce the
stabilizing effect of stable stratification via the weakening
of the buoyancy restoring force. To highlight this effect,
it is illuminating to consider the extreme
limit of strong thermal
diffusion where density fluctuation becomes stationary 
with $\p_t \rho' = N^2  v_x/g - \mu k^2 \rho'=0$ in Eq. (\ref{eq3}).
In this limit, by following a similar analysis as previously (with
$D = \nu$),
we can easily obtain the vertical and horizontal
particle diffusivities as follows:
\begin{eqnarray}
D_{xx}
&\sim & {{\tau_f} \over 2 \alpha^2}\int {d^3k \over (2 \pi)^3}
{\phi_{11}(\bk)\over (\gamma + a^2)^2}\,,
\label{eq009}\\
D_{yy}
&\sim & {{\tau_f} }\int {d^3k \over (2 \pi)^3 \gamma^2}
\left[{a^2 \phi_{11}(\bk)\over \gamma^2 (\gamma + a^2)^2}{1\over 2 \alpha^2}
+{\beta^2 \phi_{22}(\bk)\over  (2 \nu k^2)^2}\right]\,,
\label{eq010}
\end{eqnarray}
where $\alpha = \gamma N^2/ \mu k^2 (\gamma+a^2)$.
Equation (\ref{eq009}) shows that the vertical
mixing $D_{xx} \propto \mu^2/N^4 \propto Pe^{-2}$ decreases 
for large $N$ while increasing for large $\mu$. This is because
the reduction in the vertical mixing due to buoyancy force
is weaken by a strong radiative damping $\mu$. 
The comparison of Eqs. (\ref{eq009})-(\ref{eq010}) further shows that 
the reduction in vertical mixing relative to horizontal mixing
is given by a factor of
$(\nu k^2)^2/\alpha^{2}$
for an isotropic forcing and is thus weaker
than that in the case of weak 
radiative damping $\mu k^2 \ll N$ [see Eqs. (\ref{eq019})-(\ref{eq021})].
Furthermore, this reduction appears
on scales larger than the critical scale  $l_{c\mu} =
(\mu \nu)^{1/4}/N^{1/2} \sim 10 l_c$. Here,
$l_c=\sqrt{\nu/N}$ is  the
critical scale in the case of
$\mu k^2 \ll N$. These results thus show that a strong thermal diffusion
weakens the buoyancy effect and
makes the effect of stratification
become important on larger scales, compared to
the case of a weak thermal diffusion.
This result is thus consistent with the expectation
employed in previous works.


To summarize, a stable stratification can dramatically quench turbulent
transport with a more effective mixing in the horizontal directions
orthogonal to the background density gradient. It does not however
affect the amplitude of the turbulent flow,
the ratio of which is found to depend only on $\nu/\mu$
(for a temporally short-correlated forcing). 

\renewcommand{\arraystretch}{1.5}
\begin{table}[h]
\begin{tabular}{|c|c|c|c|}
\hline
& $\OOmega=0$, $N\ne 0$ & $\OOmega \ne 0$, $N=0$ & $\OOmega \ne 0$, $N\ne 0$ 
\\ \hline
$\langle \mathrm{v}_x^2 \rangle$ & $\mu^{-1}$ & 
$\OOmega^{-1}$ & $\OOmega^{-1} $ \\ \hline
$\langle \mathrm{v}_y^2 \rangle \sim \langle \mathrm{v}_z^2 \rangle$ 
& $\nu^{-1}$ & $\OOmega^{-1} \xi_\nu^{-1/3}$ & $\OOmega^{-1} 
\xi_\nu^{-1/3}$  \\ \hline
$ \nu_T $ & $0$ & $\OOmega^{-2} $ & -$\OOmega^{-2} $ \\ \hline
$D_T^{xx}$ & $N^{-2}$ & $\OOmega^{-2}$ & 
$N^{-2} \xi_D \xi_\mu^{-2/3} $ \\ \hline
$D_T^{yy} \sim D_T^{zz}$ & 
$\nu^{-1} (\nu+D)^{-1}$ &
$\OOmega^{-2} \xi_\nu^{-2/3}$ &
$\OOmega^{-2} \xi_\nu^{-2/3}$ \\ \hline
$\langle \mathrm{v}_y^2 \rangle/ \langle \mathrm{v}_x^2 \rangle$ 
& $\mu/\nu$ & $\xi_\nu^{-1/3}$ &  $\xi_\nu^{-1/3}$  \\ \hline 
$D_T^{yy}/D_T^{xx} $ & $N^2$ & $\xi_\nu^{-2/3}$ & 
$(N/\OOmega)^2 (\mu/\nu)^{2/3} \xi_\nu^{-1}$ \\ \hline
\end{tabular}
\caption{\label{Summary} Scaling of turbulence amplitude, 
turbulent viscosity ($\nu_T$), and turbulent diffusivity $D_T$ 
for an isotropic
forcing with $\phi_{11}\sim \phi_{22}$ in the case $N \gg \mu k^2$
and $D \sim \nu \ll \mu$.
$\xi_\mu = \mu k_y^2/\OOmega \ll 1$, $\xi_\nu = \nu k_y^2/\OOmega
\ll 1$, and $\xi_D = D k_y^2/\OOmega \ll 1$ are small
parameters representing strong shear limit.
The second ($\OOmega=0$ and $N\ne 0$), the third ($\OOmega \ne 0$
and  $N=0$), and the fourth ($\OOmega \ne 0$ and $N\ne 0$)
columns contain the results for stratified unsheared case in
Sec. 2, for unstratified sheared case (Kim 2005), and for
stratified sheared case in Sec. 3, respectively.
}
\end{table}

\section{Consistent theory}
The results in Sec. 2 showed that for a temporally short correlated
forcing, a stable stratification reduces turbulent transport only,
leading to anisotropic turbulent transport,
without much effect on turbulence level.
In this section, we study how these results are modified by
a stable background shear flow (differential rotation in the tachocline).
In particular, we will show that a shear flow not only 
inhibits the vertical mixing further, enhancing the anisotropic transport,
but also reduces turbulence levels anisotropically, 
thereby leading to effectively stronger
horizonal turbulence.
For simplicity, we
ignore the latitudinal differential rotation compared
with the radial differential rotation since it is weaker in the tachocline
due to thin tachocline ($h <0.03 \sim 0.05$ of the solar radius $R$).
The inclusion of the latitudinal differential rotation would introduce
a small correction term in our results. For instance, for turbulence
amplitude, this correction term is of order $(h/R)^2\ll 1$, as shown
in Leprovost and Kim (2006). Note that the latitudinal shear 
is crucial for non-vanishing horizontal momentum transport
in Leprovost \& Kim (2006).
Again, we envision that turbulence is
maintained in the tachocline by an external forcing while
gravity waves are excited due to this external forcing
in the stably stratified tachocline. We shall then compute
the overall turbulent
transport consistently by taking into account the interaction among
turbulence, shear flow and gravity waves,
instead of simply assuming a (large) constant value
of turbulent viscosity for mean shear flow (and gravity waves). 
Note that this treatment is essential 
when there is no clear scale separation between gravity waves 
and turbulence, in which case turbulence cannot be considered to give an
enhanced value of viscosity for gravity waves (c.f. Charbonnel \& Talon 2005).

For the evolution of fluctuations, we approximate
the radial differential rotation
by a linear shear flow with ${\bf U}_0 = -x \OOmega {\hat y}$
to keep the analysis tractable.
Here, $\OOmega$ is the shearing rate which we assume to be positive without
loss of generality.
As done in previous papers (Kim 2005; Leprovost \& Kim 2006), 
to capture the effect of
shearing due to radial differential rotation 
($\OOmega \sim 3 \times 10^{-6}$ s$^{-1}$ for the tachocline) 
non-perturbatively,
we use
the special Fourier transform for fluctuating quantities $\phi'$:
\begin{equation}
\phi'({\bf x},t) = {1\over (2 \pi)^3} \int d^3 k
\tilde{\phi}({\bf k},t) \exp{\{i(k_x(t) x + k_y y + k_z z)\}}\,.
\label{eq9}
\end{equation}
Here, $k_x=k_x(t)$ is 
the time dependent [unlike constant $k_x$ in Eq. (\ref{eq09})], satisfying
an eikonal equation
\begin{equation}
\p_t k_x(t) = k_y \OOmega\,.
\label{eq10}
\end{equation}
Equation (\ref{eq10}) implies that
$k_x$ linearly increases in time as $k_x(t) = k_x(0) + k_y \OOmega t$,
manifesting the main effect of shearing by a shear flow $U_0(x){\hat y}$,
i.e., generation of fine scales in the $x$ direction due to tilting and 
distortion of
fluid eddies (e.g. see Burrell 1997; Kim 2004; Kim 2005). 
The efficient generation of fine scales by shearing leads to the
breakup of eddies and enhancement of the overall dissipation,
thereby reducing turbulence amplitude and transport (e.g. see Kim 2005).
A similar effect by a shear flow is expected to persist for a more  
realistic radial shear [e.g. for an error function 
used in helioseismic inversions (Kosovichev 1996; Corbard et al 1999)]
since the basic mechanism 
of shearing (e.g. see Fig. 2 in Kim 2005) is the same 
regardless of the details of the profile of radial shear. The 
efficiency of the shearing could depend on the details of 
the profile, possibly leading to a slightly different scaling.

It is interesting to note that
a gravity wave with an initially positive value of $k_x(0)$
can change its sign after the time interval $k_x(0)/k_y \OOmega$
(for $k_y>0$) due to shearing. Since the local (radial) group velocity of
gravity waves is given by $v_{gx} = - k_y^2 k_x N^2/k^4(\omega-U_0 k_y)$
(e.g., see Kim \& MacGregor 2003),
the gravity wave thus alters its propagation direction
as $k_x$ flips its sign. Therefore, a gravity wave which initially
propagates downward to the
interior from the convection zone with a negative vertical group
velocity (i.e. $v_{gx}<0$) can propagate
upwards when the vertical group velocity becomes
positive ($v_{gx}>0$) due to shearing. 

For parameter values typical of the tachocline $N \sim
3 \times 10^{-3}$ s$^{-1}$ and $\OOmega \sim 3 \times 10^{-6}$ s$^{-1}$,
$Ri = N^2/\OOmega^2 \gg Ri_c \simeq 1/4$,
satisfying the stability criterion (Lighthill 1978).
We thus assume 
that the radial shear flow is stable 
with large value of
$Ri =N^2/\OOmega^2  \gg 1$ in the remainder of the paper (see also Schatzman, Zahn, \& Morel 2000). Note that the buildup of chemical composition gradient (the so--called
$\mu$ gradient in the solar interior 
[e.g. see Michaud and Zahn (1998) and references therein]
would further increase the values of $N$ and $Ri$, making the radial shear
flow more stable, although this effect could be counteracted by
a radiative damping, as shown in Sec 2. The shearing rate $\OOmega \sim 3 \times 10^{-6}$ s$^{-1}$ 
due to this radial differential rotation is larger than
the dissipation rate due to radiative damping $\mu k_y^2$
on a broad range of scales $l(=1/k_y) >10^{6} \sim 10^7 $ cm 
$\sim 10^{-3} H_0$, 
where $H_0$ is the pressure
scale height $\sim 6 \times 10^9$ cm. 
Thus, we focus on the strong shear limit in the following by using 
$\xi_\mu = \mu k_y^2/\OOmega
\ll 1$ as a small parameter.

For $Ri\gg 1$ and $\nu \sim D \ll \mu$,
a long but straightforward algebra can give us the solutions to 
Eqs. (\ref{eq1})-(\ref{eq4}), as shown
in Appendix A. By using these solutions and
the correlation functions of forcing defined in Eq. (\ref{forcing}), we obtain
the following results for
turbulence level
and transport coefficient defined by
$\langle n' v_i \rangle = -D_T^{ij} \p_j n_0$ for $i=1,2$ and $3$, and
momentum flux $\langle v_x v_y \rangle = -\nu_T \p_x U_0 $ 
in the strong shear limit $\xi_\mu = \mu k_y^2/\OOmega \ll 1$
(see Appendix A for details):
\begin{eqnarray}
\langle v_x^2 \rangle
&\simeq & {{\tau_f} \over \OOmega}\int {d^3k \over (2 \pi)^3}
{\phi_{11}(\bk)\over 2 \gamma^{3/2}}\,,
\label{eq119}\\
\langle v_y^2 \rangle
&\simeq & {{\tau_f} \over \OOmega}\int {d^3k \over (2 \pi)^3}
\left[{\phi_{11}(\bk)\over 2 \gamma^{5/2}} |\ln{\xi_\mu}|
+{\beta^2 \phi_{22}(\bk)\over  \gamma^2 } G_0 \right]\,,
\label{eq120}\\
\langle v_z^2 \rangle
&\simeq& {{\tau_f} \over \OOmega}\int {d^3k \over (2 \pi)^3}
\left[{\beta^2 \phi_{11}(\bk)\over 2 \gamma^{5/2}} |\ln{\xi_\mu}|
+{\phi_{22}(\bk)\over  \gamma^2} G_0 \right]\,,
\label{eq121}\\
D_T^{xx}
&\simeq& {\tau_f\over  2 N^2} \int {d^3k \over (2 \pi)^3}
{\phi_{11}(\bk)\over  \gamma \sqrt{\gamma+a^2}} \xi_D G_1\,,
\label{eq211}\\
D_T^{yy}
&\simeq& {\tau_f\over  2 } \int {d^3k \over (2 \pi)^3 \gamma^2}
\biggl[{\phi_{11}(\bk)\Gamma({4\over 3})\over 2 N^2\gamma \sqrt{\gamma + a^2}} 
{D \over \mu}
\left( {3 \over \xi_\mu}\right)^{{1\over 3}} 
+{\beta^2 \phi_{22}(\bk)\Gamma({5\over 3})\over  \OOmega^2  } 
\left( {3 \over 2 \xi_\nu}\right)^{{2\over 3}} 
\biggr]\,,
\label{eq212}\\
\nu_T
&\simeq& - {\tau_f\over  2 } \int {d^3k \over (2 \pi)^3}
{\phi_{11}(\bk)\over \gamma(\gamma+a^2)}
\left[ {1\over \OOmega^2} +
{1\over 12N^2(\gamma+a^2) }  \right]\,.
\label{eq213}
\end{eqnarray}
Here, $\xi_\nu = \nu k_y^2/\OOmega \ll 1$, $\xi_D = D k_y^2/\OOmega \ll 1$,
$\xi_\mu = \mu k_y^2/\OOmega\ll 1$,
$G_0 ={1\over 3}\Gamma\left({1/ 3}\right)\left({3/ 2\xi_\nu}\right)^{1/3}$ 
and $G_1={1\over 3} \Gamma\left({2/ 3}\right)
\left({3/ \xi_\mu}\right)^{2/3}$. Note again that $\xi_\mu \ll 1$ is valid
on a broad range of scales $l(=1/k_y) >10^{6} \sim 10^7 $ cm 
in the tachocline and also that $\xi_\mu \ll 1$ guarantees
that $\xi_D \sim \xi_\nu \ll 1$ since $D \sim \nu \ll \mu$.  The spectrum $\phi_{ij}$ in Eqs. (\ref{eq119})--(\ref{eq213}) are given
by Eqs. (\ref{forcing}), which are related to power 
spectrum of forcing $\psi_{ij}$ in Eq. (\ref{eq18}) in
the incompressible case.

Equations (\ref{eq119})-(\ref{eq213}) reveal the following interesting
features. 
Turbulence levels given in Eqs. (\ref{eq119})-(\ref{eq121}) 
are again independent of stratification, similarly to the
case without
a shear flow [Eqs. (\ref{eq19})--(\ref{eq21})] while
they 
are reduced for strong shear $\OOmega$.
This indicates that waves and shear flows play
different roles in turbulence regulation -- waves
do not necessarily quench fluctuation levels while shear flows
can reduce them through enhanced dissipation via shearing.
The turbulence regulation by shearing gives us
horizontal velocity fluctuations in Eqs. (\ref{eq120}) and
(\ref{eq121}) which are effectively higher 
than vertical one in Eq. (\ref{eq119}).
While a similar tendency was also found in
the absence of gravity waves (Kim 2005), the exact value
of the ratio of  vertical to horizontal turbulence levels
is not the same.
For example, $\langle v_x^2 \rangle/\langle v_y^2 \rangle
\propto (|\ln{\xi_\mu}|)^{-1}$ for a strong radial
forcing $\phi_{11}$ with $\phi_{22}=0$ while $\langle v_x^2 \rangle/\langle v_y^2 \rangle
\propto \xi_\nu^{1/3}$ for an isotropic forcing
with $\phi_{11} \sim \phi_{22}$. Note that 
$\xi_{\nu} \ll \xi_\mu \ll 1 $ are small parameters
in our problem, representing the strong shear limit.
These results are to be compared with
$\langle v_x^2 \rangle/\langle v_y^2 \rangle
\propto \xi_\nu^{1/3}$ in the unstratified medium (Kim 2005). The results for the isotropic forcing in various cases
are summarised and compared in Table 1.

Transport properties in Eqs. (\ref{eq211})--(\ref{eq213}) however 
exhibit a very different behaviour, with both
vertical and horizontal mixing being inhibited in a non-trivial manner
by strong stratification as well as by shearing. 
First, the vertical transport is reduced as $D_T^{xx} \propto 
\xi_D \xi_\mu^{-2/3}N^{-2}\propto D \mu^{-2/3} \OOmega^{-1/3} N^{-2}$,
becoming small as either stratification or shearing increases.
Note that the decrease in $D_T^{xx}$ for large $N$ 
agrees with Miesch (2003).
Interestingly, $D_T^{xx}\propto \mu^{-2/3} \propto Pe^{2/3} $ 
decreases as the radiative damping $\mu$
increases. 
This is because large radiative damping increases
thermal dissipation, 
thereby mainly inhibiting the vertical mixing, as noted
in Sec. 2. 
Thus, compared to the case with $N = 0$ 
where $D_{T}^{xx} \propto \OOmega^{-2}$ (Kim 2005),
the vertical mixing is much more quenched by a factor
of $\xi_D \xi_\mu^{-2/3} (\OOmega/N)^2 \ll \xi_D \xi_\mu^{-2/3}
= (D/\mu) \xi_\mu^{1/3} \ll \xi_{\mu} \ll 1$ since $\OOmega/N < 1$ 
and $D/\mu \sim 10^{-5}$ in the tachocline (see also Table 1). This clearly shows that shear flow (orthogonal to radial
density gradient), stable stratification, 
and radiative damping
can all inhibit the radial transport.

Second, $D_T^{yy}$ 
is less affected by stratification since
it involves the two parts -- the one from $\phi_{11}$
is proportional 
{to $\xi_D \xi_\mu^{-4/3}N^{-2}$ while the other
from $\phi_{22}$ is proportional to $\OOmega^{-2} \xi_\nu^{-2/3}$,
independent of $N$.
In the simplest case of a strong radial forcing
with $\phi_{22}=0$, $D_T^{xx}/D_T^{yy} \sim \xi_\mu^{2/3} \ll 1$,
independent of $N$. This is  similar to the case without stratification where
$D_T^{xx}/D_T^{yy} \sim \xi_\nu^{2/3}$ (Kim 2005).
In the general case where $\phi_{11}\ne 0$ and $\phi_{22}\ne 0$, $D_T^{yy}$ 
exhibits 	
a non-trivial, complex interplay among stratification and shear flow
in determining the overall transport. To appreciate this,
we compare Eq. (\ref{eq212}) with the 
result obtained without shear flow (\ref{eq020}) to find 
that shear flow enhances the 
contribution from $\phi_{11}$ by a factor 
$\xi_D \xi_\mu^{-4/3} = (D/\mu) \xi_\mu^{-1/3}$
while it reduces the part from $\phi_{22}$ by a factor
of $ (\nu/ \OOmega)^{2} \xi_\nu ^{-2/3}
\propto \xi_\nu ^{4/3}\ll 1$. Since $D/\mu \sim 10^{-5} \ll 1$
and $\xi_\mu \ll 1$, the horizontal transport driven by a strong
radial forcing $\phi_{11}$ can be either inhibited or enhanced
by a shear flow (due to $\phi_{11}$) depending on the parameter values.
On the other hand, the horizontal transport due to $\phi_{22}$ is reduced by the shear flow by a 
factor proportional to $\xi_\nu^{4/3} \ll 1$. 
The end result can easily be shown to be 
the enhancement of the effect of stratification.
To see this, we compare the two contributions from $\phi_{11}$
and $\phi_{22}$ to $D_T^{yy}$ and  obtain
a characteristic scale $l_*=(D/\mu)^2(\nu/N)^{1/2} (\OOmega/N)^{5/2} $
 above which the 
stratification is important, being mainly responsible for 
quenching transport. 
Since $l_*$ is smaller than $l_c=(\nu/N)^{1/2}$ obtained without a 
shear flow, the effect of stratification becomes important
compared to the case without shear flows.
It is also worth comparing the contributions to $D_T^{yy}$
from $\phi_{11}$ and $\phi_{22}$ separately
to those in the case of $N = 0$ in Kim (2005), 
where $D_T^{yy} \propto \OOmega^{-2}\xi_\nu^{-2/3}$ from both $\phi_{11}$
and $\phi_{22}$. That is, the contribution from $\phi_{11}$
is further reduced by a factor of $(\OOmega/N)^2 \xi_\nu^{1/3} (\nu/\mu)^{4/3}$ by
stratification, while the one from $\phi_{22}$ is not affected.

We now examine how the stratification affects the anisotropy
in transport. Equations (\ref{eq211}) and (\ref{eq212})
for a strong radial forcing with $\phi_{11} \gg
\phi_{22}$ and $a \sim 0$ gives 
$D_T^{xx}/D_T^{yy} \sim \xi_\mu^{2/3} \ll 1 $, independent of $N$. This is analogous to the result $D_T^{xx}/D_T^{yy} \sim \xi_\nu^{2/3} \ll 1 $
obtained in the unstratified case (Kim 2005). In this case,
the anisotropy in the transport is solely caused by 
shear stabilization.  For an isotropic forcing with 
$\phi_{11} \sim \phi_{22}$ and $a \sim 1$,
$D_T^{xx}/D_T^{yy} \sim (\OOmega/N)^2 (\nu/\mu)^{2/3} \xi_\nu \ll 1 $.
In other words, the anisotropy in turbulent transport depends on stratification, shear, and $\nu/\mu$, as clearly shown in Table 1. Since 
$(\OOmega/N)^2 \ll 1$ and $\nu/\mu \ll 1$ in the tachocline,
$D_T^{xx}/D_T^{yy}$ 
is much
smaller than $\xi_\nu^{2/3}$ obtained in the unstratified case (Kim 2005).
That is, 
the anisotropy in transport is further enhanced due
to stratification. 
We emphasize that the anisotropy in turbulent transport
is much stronger than that in turbulence amplitude, discussed previously
(see also Table 1). This result also shows the reduction in $D_T^{xx}$ for large $\mu$ and
again highlights the importance of the radiative
damping in reducing the vertical transport, thereby increasing the anisotropy in the transport. 

Finally, Eq. (\ref{eq213}) demonstrates one of the most important
effects of a stable stratification, which is to drive a system
away from a uniform rotation with a negative eddy viscosity.
Recall that in the absence
of stratification the eddy viscosity is positive in 3D while
negative in 2D limit (e.g. see Kim 2005). In contrast, the
eddy viscosity in Eq. (\ref{eq213}) is negative in both 2D ($\beta = 0$ and $\gamma = 1$) and 3D cases. This behavior was also found in recent numerical simulation 
by Miesch (2003) of a stably stratified turbulence with an 
imposed shear driven by penetrative convection,
who found anti-diffusive radial 
and diffusive latitudinal momentum transport, thereby offering
a mechanism for a proper transition from latitudinal differential
rotation in the convection zone to solid body rotation in the radiative 
interior.
Anti-diffusive momentum transport is a generic
feature of a strongly stratified medium (i.e. a geophysical
system). 
It is interesting to note that a negative viscosity
was also found in the previous work on
momentum transport due to radiatively damped
gravity waves (e.g. Kim and MacGregor 2001;2003).
In addition, the result (\ref{eq213}) shows that the (anti-diffusive)
momentum transport becomes less efficient for strong stratification
(large $N$), in agreement with Miesch (2003).

To summarise, our results show that 
the shearing
effect by radial differential rotation together with gravity waves
is an important
mechanism for turbulence regulation, leading to a weak turbulent
transport and anisotropic turbulence and transport in the tachocline. 
Furthermore, we have consistently derived
the values of turbulence level, particle mixing and momentum transport
starting from the first principle, clearly identifying the different
roles of gravity waves and shearing in transport.
For instance, in comparison with Chaboyer and Zahn (1992) [or
Spiegel and Zahn (1992)] which
start with the assumption of a strong horizontal mixing, we 
identified the source of such an anisotropic turbulence.
In particular, we have made a clear distinction between
the anisotropy in the turbulence level and turbulent transport. 

\section{Discussion and conclusions}
We have studied turbulent transport
in the stably stratified tachocline with a strong radial differential
rotation,
when turbulence is driven and maintained by a 
forcing (e.g. due to plumes penetrating from the convection zone
or due to instability).
We have assumed that both turbulence and gravity waves are
on small scales (with no 
clear scale separation between the two) and treated the interaction
among gravity waves, turbulence and shear flow consistently.
Unlike a shear flow which regulates both turbulence level
and transport, a stable stratification is shown to mainly inhibit turbulent
transport, leading to a further reduction in transport
compared to the unstratified case (Kim 2005). 
Specifically, for parameter values typical of the tachocline
($N/\OOmega \gg 1$),
particle transport due to a strong radial forcing $\phi_{11}$
(with $\phi_{22} = 0$) is 
reduced as $\xi_D \xi_\mu^{-2/3} N^{-2}$, becoming
much smaller than horizontal transport ($\propto \xi_D \xi_\mu^{-4/3} N^{-2}$) 
by a factor of $\xi_\mu^{2/3} \ll 1$. Here, $\xi_\nu = \nu k_y^2/\OOmega \ll 1$,
 $\xi_D = D k_y^2/\OOmega \ll 1$,
and
$\xi_\mu = \mu k_y^2/\OOmega \ll \xi_\nu$ are the small parameters
characterizing a strong shear limit
(see the main text for more details). Note that in this case $D_T^{xx}/D_T^{yy} \propto \xi_\mu^{2/3} \ll 1$
depends only on shearing but not on stratification, indicating 
that the anisotropy in particle transport 
is mainly governed by radial differential rotation.
A similar scaling ($\propto \xi_\nu^{2/3} \ll 1$) 
was also found in the unstratified case
(Kim 2005). However, in the case of an isotropic forcing
with $\phi_{11} \sim \phi_{22}$ and $k_x/k_y \sim 1$,
the horizontal mixing is much less reduced with
$D_T^{yy} \sim \OOmega^{-2} \xi_\nu^{-2/3}$ (with no effect
of  stratification), leading to a stronger anisotropy in transport with 
$D_T^{xx}/D_T^{yy} \sim (\OOmega/N)^2 (\nu/\mu)^{2/3} \xi_\nu 
\ll \xi_\nu^{2/3} \ll 1 $.
Note that the anisotropy becomes stronger for larger $\mu$.

Furthermore, the vertical momentum transport
was shown to be anti-diffusive with a negative eddy viscosity.
That is, small-scale
turbulence influenced by gravity waves accentuates
the gradient in a radial differential rotation rather than
makes it uniform. 
This is similar to the tendency
obtained in the case of momentum deposition by
gravity waves due to radiative
damping (Kim \& MacGregor 2001;2003).
The sharpening of the gradient of the radial
shear due to the negative viscosity
could eventually lead to time variation in
the tachocline (similarly to Kim \& MacGregor 2001)
or instability (e.g., see Petrovay 2003),
causing a rapid radial mixing 
and thus reducing the anisotropy.

Even in
the stably stratified radiative interiors of stars as well
as in the tachocline,
background turbulence has often been assumed to 
be present to enhance the value of effective viscosity.
While the clarification of the source of turbulence
responsible for such an enhanced eddy viscosity is an interesting
problem, our results demonstrate how this residual turbulence
interacts with gravity waves and shear flow, providing the
prediction for the values of eddy viscosity as well as the particle
diffusivity which depend on physical quantities like $\OOmega$ and $N$.
In particular, the radial particle transport $D_T^{xx} \sim
\xi_D \xi_\mu^{-2/3} N^{-2}$ indicates that turbulent transport of particles
can be inhibited due to stable stratification, shear flow and 
large radiative damping.
This finding can have interesting implications for the surface
depletion of lithium in the Sun and other stars (i.e., Pinsonneault 1997),
and will be studied in a future publication. 
Furthermore, if a similar physical process operates in the bulk of the radiative
interior of the sun, a negative viscosity that we obtain implies that a radial 
differential rotation
which is created during its spin-down would not
be eliminated.
Note however that 
Charbonnel \& Talon (2005) have shown an efficient
momentum transport in the solar radiative interior due to the
cumulative effect of large scale meridional flow, shear instability and
gravity waves.

The tachocline is believed to possess a strong toroidal magnetic
fields of strength $10^4 \sim 10^5$ G, which could have
an important influence on turbulent transport in that region. 
This is an important problem since the presence of a weak
poloidal magnetic field in the radiative interior
together with a strong toroidal magnetic field in the tachocline
(acting as a boundary layer between the radiative interior and
convection zone) could offer a mechanism for a uniform rotation
in the interior as well as for the tachocline confinement 
(e.g. R\"udiger \& Kichatinov 1996; MacGregor \& Charbonneauu 1997;
Gough \& McIntyre 1998) 
[see however Brun \& Zahn (2006) for a negative
result on this scenario].
In this case, the values of effective diffusivity of magnetic
fields as well as eddy viscosity play a crucial role in determining
the thickness of the tachocline. Therefore, a consistent computation
of magnetic diffusivity and eddy viscosity with their
dependences on physical quantities such as the strength of magnetic fields, 
the Brunt-V\"ais\"al\"a frequency, 
and shearing rate, would be of primary interest 
(Kim and Leprovost 2007).
It is an interesting question, in general, whether a negative eddy viscosity,
favored in a stably stratified medium, remains as a robust feature in
the presence of magnetic field in view of the forward
energy cascade (i.e. positive eddy viscosity) in MHD turbulence.
Furthermore, composition gradients (discussed in Sec. 3)
and meridional flows
would contribute to the transport in the tachocline. In particular,
meridional flows are expected to enhance the radial transport
of chemical species by advection as discussed in Kim (2005)
although this enhancement would be reduced for stronger
horizontal turbulent [e.g., see Kim (2005) and Chaboyer and Zahh (1992)].
These issues are however outside the scope of this paper and will be
addressed in future publication.

\begin{acknowledgements}
This work was supported by the UK PPARC grant PP/B501512/1.
\end{acknowledgements}
 
\appendix

\section{}

In this Appendix, we show how to derive
Eqs.\ (\ref{eq119})--(\ref{eq213}).  To this end, 
we use
$U_0 (x) {\hat y} = - x \OOmega{\hat y}$
in Eqs.\ (\ref{eq1})--(\ref{eq4}), 
and introduce the 
transforms ${\hat w}$ and ${\hat {\hat w}}$
for $w=v_i,n,p, h_i$ and $f$ as follows:
\begin{eqnarray}
{\hat w} &\equiv &{\tilde w} \exp{[\nu (k_x^3/3 k_y \OOmega + k_H^2 t)]}\,,
\nonumber \\
{\hat {\hat w}} &\equiv &{\tilde w} \exp{[\mu (k_x^3/3 k_y \OOmega + k_H^2 t)/2]}\,,
\label{a6}
\end{eqnarray}
where $k_H^2 = k_y^2 + k_z^2$.
We then use a new time variable $\tau = k_x/k_y + \OOmega t$ to 
rewrite Eqs. (\ref{eq1})--(\ref{eq4}) as:
\begin{eqnarray}
\OOmega \p_\tau \hv_x &=& -i \tau k_y {\hat p} -g \hrho   + \hf_x \,,
\label{a7}\\
\OOmega \p_\tau \hv_y -\OOmega \hv_x &=& -i k_y {\hat p}  + \hf_y\,,
\label{a8}\\
\OOmega \p_\tau \hv_z  &=& -i k_z {\hat p}  + \hf_z\,,
\label{a9}\\
0&=& \tau \hv_x +  \hv_y + {k_z \over k_y} \hv_z \,,
\label{a10}\\
\OOmega [\p_\tau + \xi_\mu (\gamma + \tau^2)] \trho &=& {N^2\over g} \tv_x\,,
\label{a100} \\
\OOmega [\p_\tau  +\xi_D (\gamma + \tau^2)] \tn &=& -(\p_x n_0) \tv_x\,.
\label{a101}
\end{eqnarray}
For a strong shear limit where $\mu = \mu k_y^2/\OOmega \ll 1$,
Eqs. (\ref{a7})--(\ref{a100}) can be combined to yield
the equation for $\hhrho$ as
\begin{equation}
\p_\tau[(\gamma +\tau^2)\p_\tau \hhrho] + {\gamma N^2\over \OOmega^2}
 \hhrho \simeq {N^2\over g \OOmega^2} {\hhh_1}(\tau)\,,
\label{a102}
\end{equation}
to leading order in $\xi_\mu = \mu k_y^2/\OOmega \ll 1$.
Here, $\hhh_1(\tau) =  \gamma \hhf_x - \tau \hhf_y - \tau \beta \hhf_z$;
$\gamma = 1+\beta^2$ and $\beta=k_z/k_y$.
In the limit of $N^2/\OOmega^2 \gg 1$, Eq. (\ref{a102})
can be solved with the following solutions valid up to O$(\mN^{-2})$:
\begin{eqnarray}
\hhrho(\tau) &=&
{N_*^2\over  \mN\OOmega}{1\over (\gamma+\tau^2)^{1/4}}
\int ^\tau {d\tau_1  \over (\gamma + \tau_1^2)^{1/4}}
\sin{{ [\vphi(\tau)-\vphi(\tau_1)]}} \hhh_1(\tau_1) \,,
\nonumber \\
\hhv_x(\tau) &=&
{1\over \mN\OOmega}{1\over (\gamma+\tau^2)^{5/4}}
\int ^\tau {d\tau_1  \over \vpsi(\tau_1) (\gamma + \tau_1^2)^{1/4}}
\biggl[ -{\tau\over 2} \sin{{ [\vphi(\tau)-\vphi(\tau_1)]}}
\nonumber \\
&&
+\mN \sqrt{\gamma+\tau^2} \vpsi(\tau) \cos{{[\vphi(\tau)-\vphi(\tau_1)]}}
\biggr]
\hhh_1(\tau_1) \,,
\nonumber \\
\hhv_y(\tau) &=&
-{\tau \over \gamma} \hhv_x(\tau) + {\beta^2 \over \OOmega \gamma} {\hhrho}(\tau)
- {\beta\over \gamma \OOmega} \int^\tau d\tau_1 
{\overline G}(\tau, \tau_1) \hhh_2(\tau_1)\,,
\nonumber \\
\hhv_z(\tau) &=&
-{\tau \beta \over \gamma} \hhv_x(\tau) - {\beta \over \OOmega \gamma} {\hhrho}(\tau)
+ {1\over \gamma \OOmega} \int^\tau d\tau_1 
{\overline G}(\tau, \tau_1) \hhh_2(\tau_1)\,.
\label{a13}
\end{eqnarray}
Here, again $\beta=k_z/k_y$, $\gamma = 1 + \beta^2$;
$N_*^2 = N^2/g\OOmega$, $\mN^2 = N^2 \gamma/\OOmega^2$;
${\overline G}(\tau, \tau_1) = \exp{\{({\xi_\mu/ 2}-\xi_\nu)
[(\gamma \tau + \tau^3/3)-(\gamma \tau_1 + \tau_1^3/3)]\}}$;
$\xi_\nu = \nu k_y^2/\OOmega$;
\begin{eqnarray}
\hhh_1&= & (1+\beta^2)\hhf_x - \tau_1 \hhf_y - \tau_1 \beta \hhf_z\,,
\nonumber \\
\hhh_2&= & - \beta \hhf_y + \hhf_z\,,
\nonumber \\
\vphi(\tau)&=& {\mN \alpha \over 2}
\ln{{\sqrt{\gamma+\tau^2}+\tau \over \sqrt{\gamma+\tau^2}-\tau}}
-{1\over 8 {\mN}}{\tau\over \sqrt{\gamma+\tau^2}}\,,
\nonumber \\
\vpsi(\tau)&=&
\alpha - {1\over 8 \mN^2} {\gamma \over \gamma+\tau^2}\,,
\nonumber \\
\alpha &= &1 - {1\over 8 \mN^2}\,.
\label{aaa}
\end{eqnarray}
By following a similar algebra used in Kim (2005) and by
using the correlation function of the forcing given in Eq. (\ref{forcing}),
we can obtain the following correlation fictions
$\langle v_i v_j \rangle$ to leading
orders in $1/\mN^2\ll 1$ and $\xi_\mu = \mu k_y^2/\OOmega\ll 1$:

\begin{eqnarray}
\langle v_x^2 \rangle
&\sim&  {\tau_f\over 2 \OOmega}
\int {d^3k d\tau\over (2\pi)^3}
{\phi_{11}(\bk) \over \sqrt{\gamma +  a^2}}
{e^{-\xi_\mu Q(\tau,a)}\over (\gamma + \tau^2)^{3/2}}
\left[ 1 + \cos{{[2 (\vphi(\tau)-\vphi(a))]}}\right]\,,
\nonumber \\
\langle v_y^2 \rangle
&\sim &  {\tau_f\over  \OOmega}
\int {d^3k d\tau \over (2\pi)^3}
\biggl[e^{-\xi_\mu Q(\tau,a)} 
{\phi_{11}(\bk) \over 2\sqrt{\gamma +  a^2}}
{\tau^2\over \gamma^2 (\gamma+\tau^2)^{3/2}}
+ e^{-2\xi_\nu  Q(\tau,a)}{\beta^2 \phi_{22}(\bk)\over \gamma^2} \biggr]\,,
\nonumber \\
\langle v_z^2 \rangle
&\sim &  {\tau_f\over  \OOmega}
\int {d^3k d\tau \over (2\pi)^3}
\biggl[e^{-\xi_\mu Q(\tau,a)} 
{\phi_{11}(\bk) \over 2\sqrt{\gamma +  a^2}}
{\tau^2\beta^2 \over \gamma^2 (\gamma+\tau^2)^{3/2}}
+ e^{-2\xi_\nu  Q(\tau,a)} 
{\phi_{22}(\bk)\over \gamma^2} \biggr]\,,
\nonumber \\
\langle v_x v_y \rangle
&\sim &  -{\tau_f\over \OOmega\gamma \mN^2}
\int {d^3k d\tau \over (2\pi)^3}
{\phi_{11}(\bk) \over \sqrt{\gamma +  a^2}\vpsi(a)^2}
{e^{-\xi_\mu Q(\tau,a)} \tau\over (\gamma+\tau^2)^{5/2}}
\nonumber \\
&&\times \left[ -{\tau\over 2} \sin{{[\vphi(\tau)-\vphi(a)]}}
+\mN \sqrt{\gamma+\tau^2} \vpsi(\tau) \cos{{[\vphi(\tau)-\vphi(a)]}}
\right]^2 \,.
\nonumber \\
&&
\label{a17}
\end{eqnarray}
Here, $a = k_x/k_y$, $Q(\tau, a) = (\gamma \tau+ \tau^3/3)-
(\gamma a +a^3/3)$;
special symmetries $\psi_{ij} (k_y) = \psi_{ij}(-k_y)$
and $\psi_{ij} (k_z) = \psi_{ij}(-k_z)$ were used.
We then compute $\tau$ integrals in Eq. (\ref{a17})
in the strong shear limit $\xi_\nu \ll \xi_\mu \ll 1$ and use
$\langle v_x v_y \rangle = - \nu_T^{xx} \p_x U_0
= \nu_T^{xx}  \OOmega$ to obtain Eqs. (\ref{eq119})-(\ref{eq121})
and (\ref{eq213}).

Next, to compute particle transport, we integrate Eq. (\ref{a101})
in time to
obtain
\begin{equation}
\tn(\bk(t),t)
= - \p_i n_0 \int dt_1 d^3 k_1
{\hat g}(\bk,t;\bk_1,t_1) e^{-DQ(t,t_1)}
{\tv}_i(\bk_1,x, t_1)\,.
\label{a18}
\end{equation}
Here, $Q(t,t_1) =  \int^t_{t_1} dt' [k_x^2(t') + k_H^2] =
[k_x^3-k_{1x}^3] /3k_y\OOmega + k_H^2 (t-t_1)$;
$k_H^2 = k_y^2 + k_z^2$ is the amplitude of wave number
in the horizontal plane; $k^2=k_H^2+ k_x^2$;
${\hat g}$ is the Green's function given by
\begin{equation}
{\hat g}(\bk,t;\bk_1,t_1)
= \delta(k_y-k_{1y})
\delta (k_z-k_{1z})
\delta
\left[k_{x} -k_{1x} - k_{1y} (t-t_1) \OOmega\right]\,.
\label{a20}
\end{equation}
A similar analysis using Eqs. (\ref{a13}), (\ref{a20}),
(\ref{eq16})--(\ref{eq18}), and (\ref{forcing}) and
$\langle n v_i \rangle = -D_{ij} \p_j n_0$ gives
\begin{eqnarray}
D_T^{xx}
&=&  {\tau_f\over 2 (2\pi)^3\OOmega^2 \mN^2}
\int d^3k d\tau
{\xi_D \phi_{11}(\bk)e^{-\xi_\mu Q(\tau,a)} \over \sqrt{\gamma +  a^2}}
{(\gamma + \tau^2)^{1/2}}\,,
\nonumber \\
D_T^{yy}
&=&  {\tau_f\over (2\pi)^3\OOmega^2}
\int d^3k d\tau {\xi_D\over \gamma^2} 
\left[e^{-\xi_\mu Q(\tau,a)} {\phi_{11}(\bk) \tau^2 \sqrt{\gamma + \tau^2}
\over 2 \mN^2 \sqrt{\gamma +  a^2}}
+ e^{-2\xi_\nu Q(\tau,a)}
\beta^2 \phi_{22} (\tau-a)^2(\gamma+\tau^2)\right]\,.
\nonumber \\
&&
\label{a22}
\end{eqnarray}
Here, $\xi_D = D k_y^2/\OOmega\sim \xi_\nu \ll 1 $.
Finally, the evaluation of $\tau$ integrals in Eq. (\ref{a22})
gives us Eqs. (\ref{eq211})--(\ref{eq212}) in the main text.

\end{document}